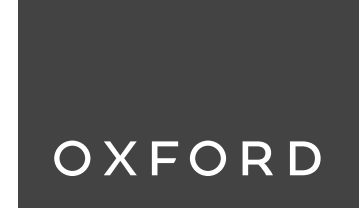

Research and Applications

# Hillclimb-Causal Inference: a data-driven approach to identify causal pathways among parental behaviors, genetic risk, and externalizing behaviors in children

**Mengman Wei, PhD[1,*] and Qian Peng, PhD[1,*]**

[1]Department of Neuroscience, The Scripps Research Institute, San Diego, CA 92037, United States

*Corresponding authors: Mengman Wei, PhD, Department of Neuroscience, The Scripps Research Institute, 10550 N Torrey Pines Rd, La Jolla, CA 92037, USA (mwei@scripps.edu) and Qian Peng, PhD, Department of Neuroscience, The Scripps Research Institute, 10550 N Torrey Pines Rd, La Jolla, CA 92037, USA (qpeng@scripps.edu)

## Abstract

**Objectives:** Externalizing behaviors in children, such as aggression, hyperactivity, and defiance, are influenced by complex interplays between genetic predispositions and environmental factors, particularly parental behaviors. Unraveling these intricate causal relationships can benefit from the use of robust data-driven methods.

**Materials and Methods:** We developed "Hillclimb-Causal Inference," a causal discovery approach that integrates the Hill Climb Search algorithm with a customized Linear Gaussian Bayesian Information Criterion (BIC). This method was applied to data from the Adolescent Brain Cognitive Development (ABCD) Study, which included parental behavior assessments, children's genotypes, and externalizing behavior measures. We performed dimensionality reduction to address multicollinearity among parental behaviors and assessed children's genetic risk for externalizing disorders using polygenic risk scores (PRS), which were computed based on GWAS summary statistics from independent cohorts. Once the causal pathways were identified, we employed structural equation modeling (SEM) to quantify the relationships within the model.

**Results:** We identified prominent causal pathways linking parental behaviors to children's externalizing outcomes. Parental alcohol misuse and broader behavioral issues exhibited notably stronger direct effects (0.33 and 0.20, respectively) compared to children's PRS (0.07). Moreover, when considering both direct and indirect paths, parental substance misuse (alcohol, drugs, and tobacco) collectively resulted in a total effect exceeding 1.1 on externalizing behaviors. Bootstrap and sensitivity analyses further validated the robustness of these findings.

**Discussion and Conclusion:** Parental behaviors exert larger effects on children's externalizing outcomes than genetic risk, suggesting potential targets for prevention and intervention. The Hillclimb-Causal framework provides a general, data-driven way to map causal pathways in developmental psychiatry and related domains.

## Introduction

Externalizing behaviors are actions directed outwardly, often disrupting others. They are commonly observed in disorders such as attention-deficit hyperactivity disorder (ADHD), conduct disorder (CD), and oppositional defiant disorder (ODD). These behaviors typically include impulsivity, hyperactivity, defiance, rule-breaking, conduct problems, as well as verbal or physical aggression, lying, and stealing, among others. These behaviors are generally considered disruptive and can negatively impact social and familial relationships.[1–3]

Youth externalizing behaviors may evolve into serious antisocial issues and adverse psychosocial outcomes in adulthood.[3] Therefore, it is crucial to study how these behaviors emerge and develop during adolescence to better understand their trajectory and inform potential interventions. Research has shown that externalizing behaviors have a genetic component, with twin studies indicating that such behaviors may be substantially heritable.[4–6] At the same time, the family environment, particularly parental behaviors, also plays a significant role in affecting children's externalizing outcomes.[3,7] The relationship between genetic predispositions and environmental influences is complex and multifaceted, warranting further investigation to better understand their combined effects.

To gain a deeper understanding of how these factors interplay and contribute to the development of externalizing behaviors in children, we developed a data-driven causal model using the Hill Climb Search algorithm.[8,9] This analysis utilized data from the Adolescent Brain Cognitive Development (ABCD) Study.[10] The ABCD Study is a longitudinal, nationally representative study of nearly 12 000 U.S. children recruited at ages 9-10 from 21 research sites. The cohort was designed to reflect the diversity of the U.S. population in terms of demographic and socioeconomic factors. Data were collected from both children and their parents or guardians, depending on the domain of interest (eg, behavioral assessments from parents, cognitive tasks from children).[10]

By applying this model, we aim to examine the interactions between parental behaviors, children's genetic predispositions, and their externalizing outcomes. The goal is to identify potential causal pathways that lead to the development of externalizing behaviors in adolescents, offering insights into early intervention and prevention strategies.

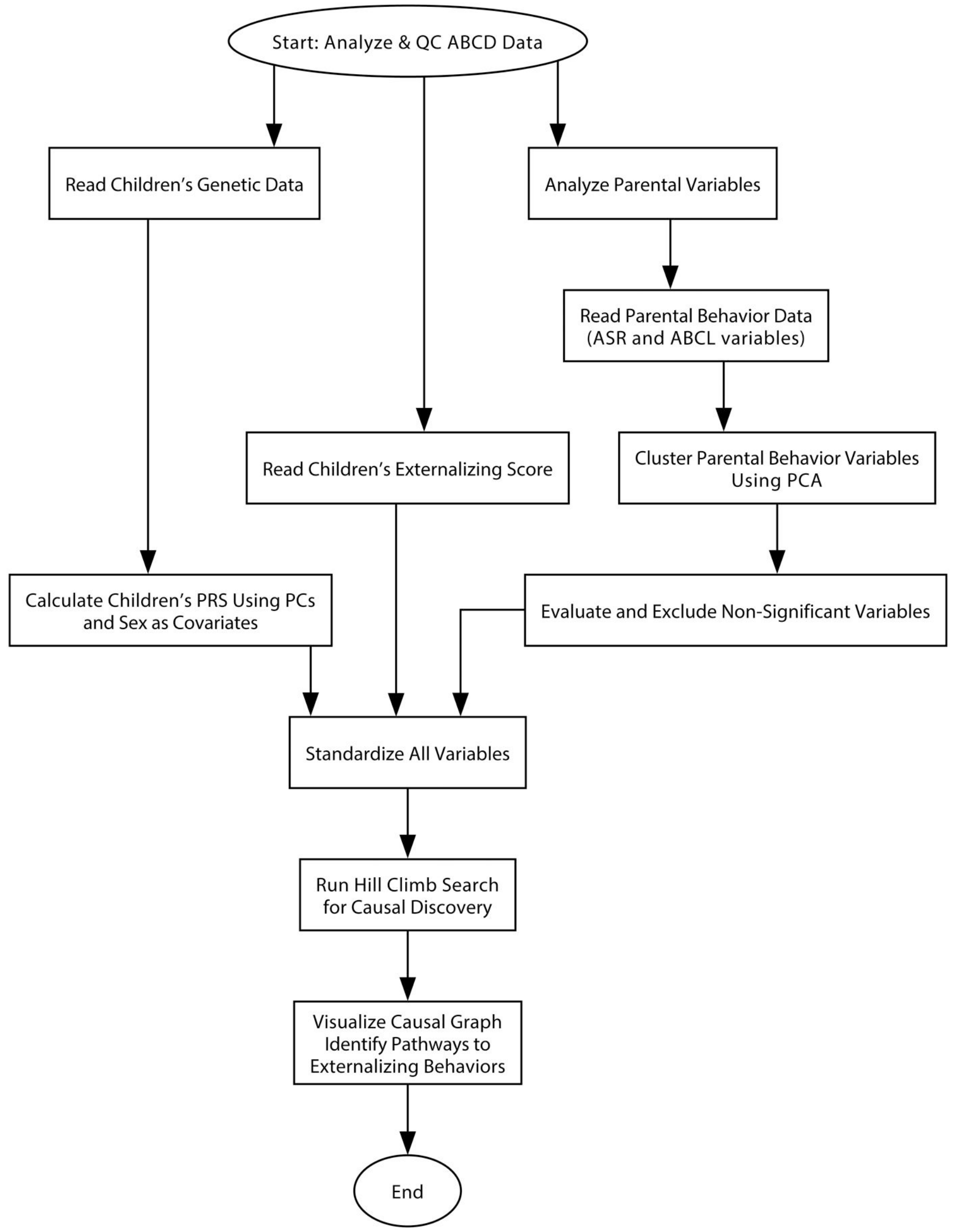

**Figure 1.** Flowchart for the whole calculation process.

## Methods

Figure 1 presents the flowchart of our study. The model takes carefully selected parental behavior factors, children's polygenic risk scores (PRS), and children's externalizing behavior T-scores as inputs. These data were preprocessed and standardized before being integrated into a hill-climb search-based causal discovery framework. This method identifies potential causal relationships among variables using data-driven associations alone. Subsequently, it outputs valid causal pairs and traces the pathways leading to externalizing behaviors using a breadth-first search algorithm. Further methodological details are described below.

### Selected parental behaviors factors

We began by examining all parental behavior T-score variables available from the Adult Self Report (ASR) and Adult Behavior Checklist (ABCL) instruments in the ABCD dataset.[11] First, we created a correlation matrix to assess how these variables were related and to check for any strong overlaps that might cause multicollinearity issues.

Next, we explored regularized regressions (Lasso, Elastic-Net, and Ridge regression) for feature selection. These methods identify the most significant features with the goal of narrowing down the number of predictors. While regularized regression is effective at identifying strong predictors, it may exclude true causal variables by penalizing and dropping correlated features. For instance, using an absolute coefficient cutoff $>0.03$ led to the exclusion of theoretically relevant variables such as parental internalizing issues. To avoid discarding important predictors prematurely, we conducted a follow-up multivariate regression analysis. This provided additional validation and allowed us to re-evaluate variables excluded during the regularization step, retaining those that met predefined criteria of statistical significance ($P$-value $<0.05$) and relative importance (LMG $>0.01$).

Finally, to address observed multicollinearity and potential interactions between selected variables, we applied hierarchical clustering and Principal Component Analysis (PCA) as dimensionality reduction techniques. These methods enabled us to identify and retain the most informative and non-redundant factors, reducing the number of variables while preserving the underlying structure, key sources of variation, and, importantly, interpretability.

### Children's genetic factor: PRS calculations

To quantify children's genetic predispositions, we calculated their PRS for externalizing behaviors using genotype data from the ABCD Study and GWAS summary statistics from the International Externalizing Consortium, excluding 23andMe data.[12,13]

### Data preparation and quality control

Prior to PRS calculation, we conducted comprehensive quality control (QC) on the ABCD genotype data. Full details of the QC procedures are provided in the Supplementary Materials. In brief, we filtered out individuals and SNPs based on standard genotype QC metrics (eg, call rate, heterozygosity, minor allele frequency, and Hardy–Weinberg equilibrium). We then performed PCA on the genetic relationship matrix to assess population stratification. The top 10 genetic principal components (PCs) were extracted and included as covariates in downstream analyses to account for ancestral diversity.

### GWAS summary statistics processing

The GWAS summary statistics for externalizing behavior were sourced from Linnér et al,[12,13] originally aligned to GRCh37 coordinates.[14] Because the ABCD genotype data are based on GRCh38,[15] we performed coordinate conversion (liftover) from GRCh37 to GRCh38 using UCSC tools.[16] Some SNPs were lost in this process due to mapping issues, which may have modestly affected PRS calculations. Additional details are provided in the Supplementary Materials.

### Polygenic risk score calculation

PRS was computed using PRSice v2.3.5,[17,18] a tool designed for high-throughput PRS analysis. We used the QC'ed ABCD genotype data and GWAS summary statistics (filtered to SNPs present in both datasets) as input. To evaluate the robustness of PRS associations, we conducted permutation testing with 10 720 permutations (equal to the number of individuals), yielding empirical *P*-values and helping to avoid overfitting. The PRS regression model included sex (coded as 0 = male, 1 = female) and the top 10 PCs as covariates. SNP-level PRSice outputs were retained for downstream analyses.

### Final analytic sample

From the initial ABCD sample of 11 880 individuals, 11 663 remained after genotype quality control. Following additional filtering steps, the final sample for PRS calculation consisted of 10 720 individuals (5621 males; 5099 females). We used genetic principal components to infer genetic ancestry and grouped individuals based on genetic similarity to reference panels. The inferred ancestry composition included individuals most similar to European (4821), African (1400), Admixed American (1220), East Asian (297), and Other/Mixed/Unassigned (2982) reference populations. For sensitivity analyses aimed at minimizing population stratification, we restricted the sample to individuals genetically similar to the European reference panel, yielding a final analytic subgroup of 4620 individuals (2416 males; 2204 females).

### Quantification of children's externalizing behaviors

Externalizing behaviors in children, encompassing symptoms and behaviors such as hyperactivity, impulsivity, aggression, oppositional behaviors, and conduct issues, were quantitatively assessed using the DSM-5-oriented T-scores available for each participant in the ABCD dataset.[19] The DSM-5 T-scores provide standardized measures of behavior severity, relative to established normative data, thus facilitating the identification and quantification of externalizing problems in children.

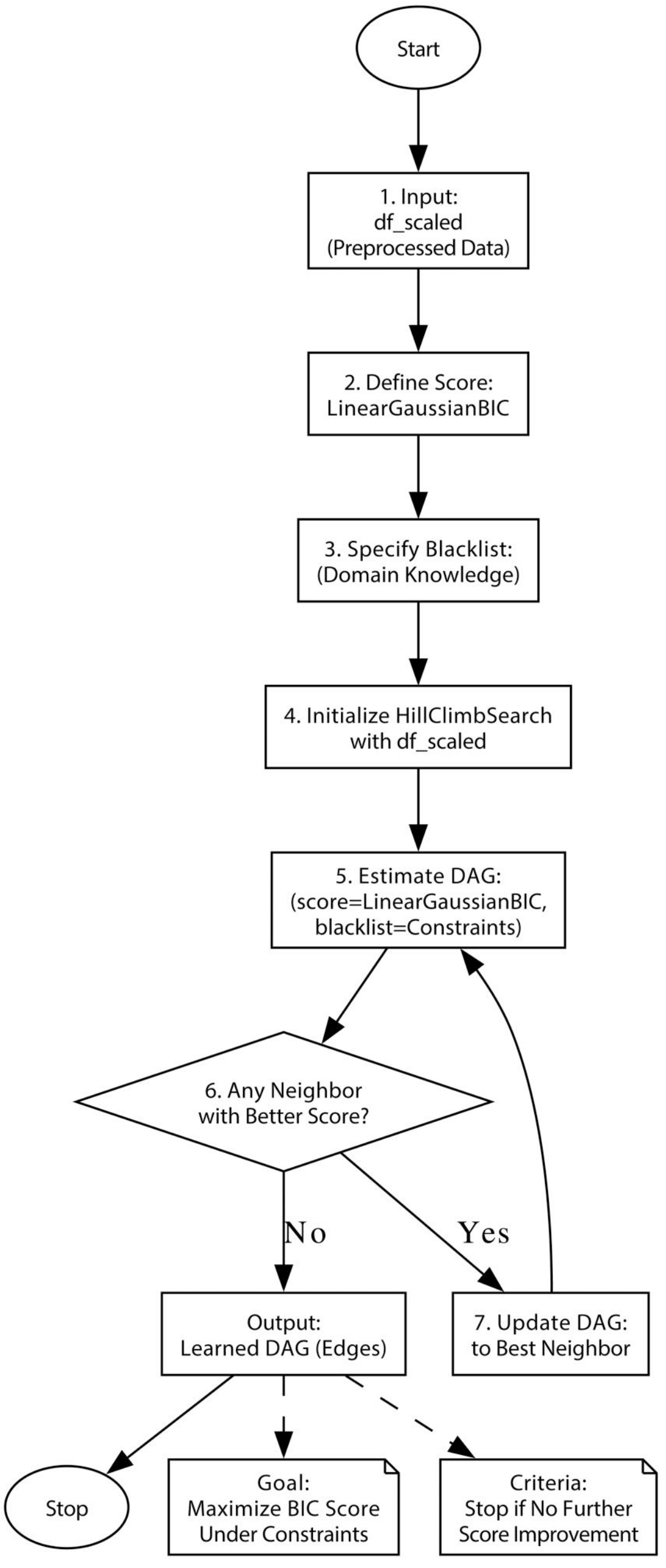

**Figure 2.** The flowchart of Hill Climb Search model.

### Hill climb search causal model

As shown in Figure 2, we applied a Hill Climb Search algorithm to learn the underlying causal structure from standardized data, representing it as a directed acyclic graph (DAG). The input included parental behavior factors, children's externalizing scores, and PRS.

The process begins by defining a scoring function and incorporating domain knowledge constraints. The algorithm then iteratively searches for improved DAG structures by evaluating neighboring graphs, continuing until no further improvement is found. The final output is a DAG that reflects the potential causal relationships among variables.

### Data preparation

We combined data from the 2-year follow-up of the ABCD study, including parental behavior measures (ASR and ABCL), children's externalizing scores (CBCL), and children's PRS for externalizing behaviors. We removed participants with missing values in any key variable, resulting in a complete dataset for analysis.

### Standardization

All continuous variables, including aggregated parental factors, PRS, and the child's externalizing behavior score, were standardized using Z-score normalization to facilitate model convergence and interpretation.

### Causal discovery using hill climb search

Causal structure learning was performed using the Hill Climb Search algorithm implemented in pgmpy.[8,9] This heuristic algorithm constructs a DAG by iteratively evaluating single-edge modifications (additions, removals, reversals) to maximize a given scoring function until a local optimum is reached. The algorithm was initialized with an empty graph, meaning no edges were present initially.

### Scoring function: Linear gaussian bayesian information criterion (BIC)

We employed a custom linear Gaussian BIC scoring method[19] appropriate for continuous variables. This score is defined as follows:

For each node $X_i$ with a set of parent nodes $\mathrm{Pa}(X_i)$ (ie, nodes with directed edges pointing to $X_i$), the local BIC score is computed by fitting a linear regression model:

$$X_i = \beta_0 + \sum_{X_j \in \mathrm{Pa}(X_i)} \beta_j X_j + \varepsilon$$

where $\varepsilon \sim \mathcal{N}(0, \sigma^2)$. The BIC for node $X_i$ given its parent set $\mathrm{Pa}(X_i)$ is:

$$\mathrm{BIC}(X_i, \mathrm{Pa}(X_i)) = -\frac{n}{2}\left[\ln(2\pi) + 1 + \ln\left(\frac{\mathrm{RSS}}{n}\right)\right] - \frac{k+2}{2}\ln(n)$$

where

1) **RSS** (residual sum of squares) measures model fit;
2) **n** denotes the sample size;
3) **k** denotes the number of predictors (ie, parent nodes), given by $|\mathrm{Pa}(X_i)|$.

The total BIC score for a DAG structure is the sum of local BIC scores across all nodes.

### Expert knowledge constraints

We imposed two logical constraints ("blacklists") to ensure biologically plausible causal structures:

The child's externalizing score (ExternalScore) was restricted from influencing other variables, as it is an outcome variable.

The PRS variable was constrained from affecting the aggregated parental behavioral factors, given the temporal ordering and the genetic basis.

### Quantification of the causal model via structural equation modeling (SEM)

The final DAG structure can be visualized using NetworkX.[20,21] Since a causal model (represented by directional arrows and potential pathways) alone does not quantify the strength of these relationships, we employed Structural Equation Modeling (SEM) to estimate both direct and indirect effects. In doing so, we aimed to verify and refine the hypothesized causal pathways leading to externalizing behaviors in children. SEM was conducted using lavaan[22] and semPlot[23] packages in R 4.3.0.[24] We specified a path model based on the structure obtained from our causal discovery, estimating both the direct and indirect effects of the parental behavior factors and PRS on externalizing behavior.

### Bootstrap analysis and sensitivity analysis

To evaluate the robustness of our causal model, we performed bootstrap and sensitivity analyses (details in the Supplementary). For the bootstrap, we generated 100 samples with replacement, refitted the DAG each time, and for each edge recorded its appearance frequency and a 95% confidence interval of its coefficient (the 2.5th and 97.5th percentiles); we report only edges with frequency $\geq 0.40$ and consider an effect consistent if the original SEM coefficient falls within the bootstrap interval. To assess generalizability, we repeated the full analytic pipeline (clustering, PCA, and DAG estimation) in the European-ancestry subset, since the PRS was derived from primarily European-based GWAS.

## Results

### Selected parental behaviors factors

The correlation matrix of all parental factors is shown in Figure 3. Features selected for further analysis are marked in bold in Table 1. For clarity, we assigned concise short names to each variable, with the original variable names and descriptive labels included in the table for reference.

The dimensionality reduction of the parental variables yielded seven aggregated factors, each capturing distinct domains while effectively mitigating multicollinearity. As illustrated in Figure 4, the "Parents Alcohol/General Substance Use" cluster, comprising Alcohol Use (ABCL) and Mean SubUse (ABCL), produced a first principal component that explained 91.6% of the variance. In contrast, the "Parents Drug Use" and "Parents Tobacco Use" clusters were retained as single variables, as their correlations with other measures did not exceed the threshold. The "Parents Overall Externalizing/Internalizing Issues (Spouse Report)" cluster grouped Aggressive, Rule-Break, Externalizing,

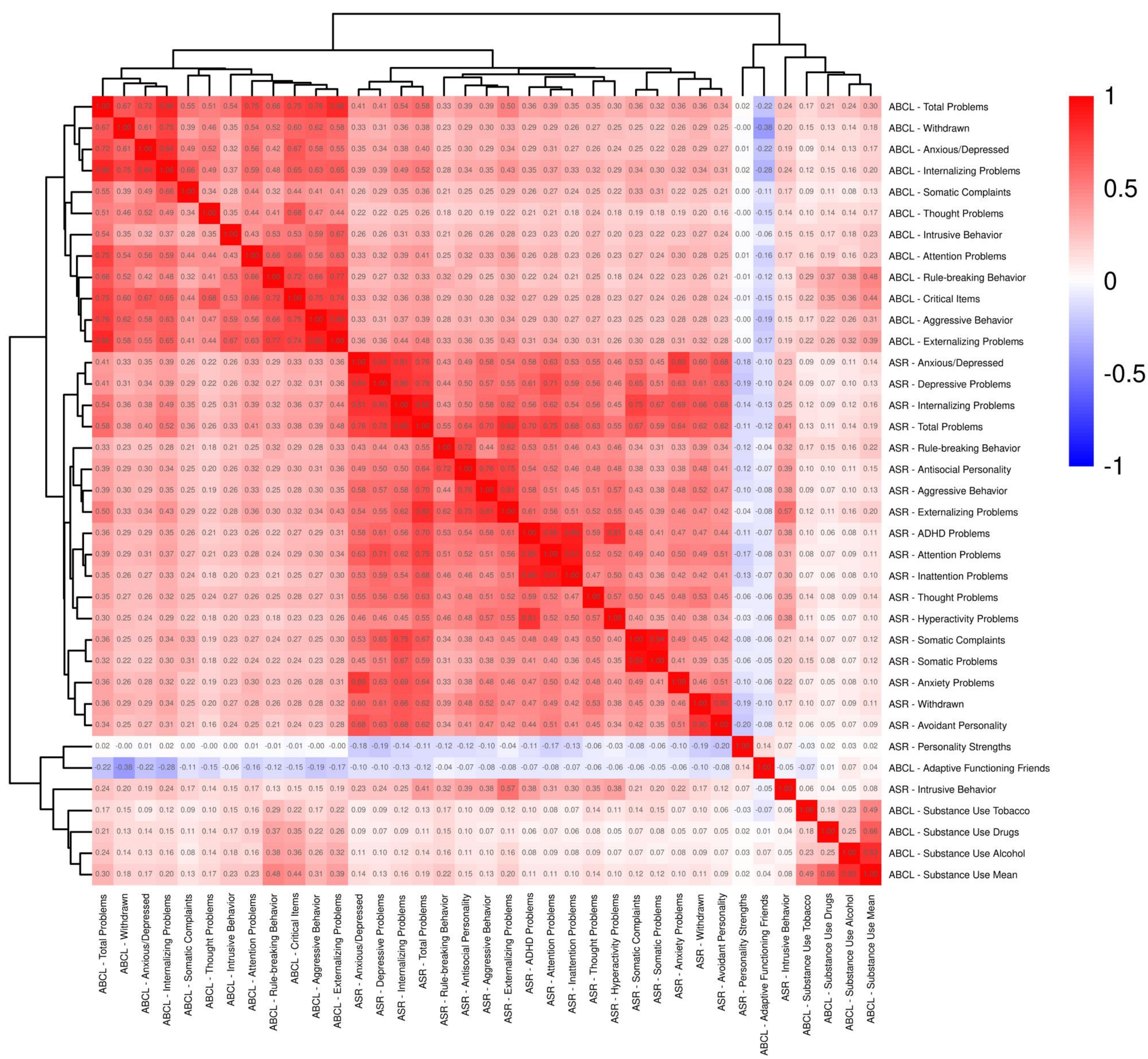

**Figure 3.** Correlation matrix for all parental behavior T-scores from ASR and ABCL lists. The numbers in each cell represent the correlation coefficients.

Internalizing, and Total Problems (all from ABCL), with its first principal component explaining 81.0% of the variance and showing the highest loadings for Externalizing, Total Problems, and Internalizing. The largest cluster, "Parents Broad Behavioral & Emotional Dysregulation (Self Report)", included 13 measures from the ASR—Aggressive, Antisocial, Avoidant, Anx/Dep, Depression, Rule-Break, Attention, ADHD, Inattention, Hyperactive, Externalizing, Internalizing, and Total Problems—with its first principal component accounting for 70.1% of the variance; here, Total Problems, Internalizing, and Externalizing had the highest loadings. Finally, the "Parents Somatic Complaints" and "Parents Intrusive Behaviors" clusters were maintained as individual variables, represented by Somatic (ASR) and Intrusive (ASR), respectively, due to their isolation based on the correlation threshold. This approach reduced the size of the original feature set while preserving the interpretability of parental behavior constructs, as demonstrated by the high variance explained by each component. The PC loadings for each aggregated factor are presented in Figure 4, illustrating the contribution of each original variable to its respective component. Overall, this approach allowed us to efficiently condense the parental behavior measures while preserving the essential variance within the data, facilitating a more robust investigation of their association with the outcomes of interest.

## Model results

The main findings from our model are presented in Figure 5. The SEM allowed us to explore both direct and indirect pathways, confirming that the significant relationships (most with $P < 0.0005$) aligned with the hypothesized causal structure identified through the Hill Climb search algorithm.

Our findings reveal three primary patterns. First, parental alcohol, tobacco, and substance-use factors are interrelated and influence one another, collectively contributing to broader parental behavioral issues. Second, among these influences, parental alcohol misuse (node `Alcohol`) exerts the strongest direct impact on children's externalizing

**Table 1.** Summary and renaming of parental behavior variables from ASR and ABCL (T-scores).

| Short name | Original variable name | Descriptive label (ABCD dataset) | Source |
|---|---|---|---|
| ASR_Perseverative | asr_scr_perstr_t | Perseverative/Obsessive Problems (T-score) | ASR |
| **ASR_AnxiousDepressed** | **asr_scr_anxdep_t** | **Anxious/Depressed Syndrome Scale (T-score)** | **ASR** |
| ASR_Withdrawn | asr_scr_withdrawn_t | Withdrawn Syndrome Scale (T-score) | ASR |
| ASR_SomaticComplaints | asr_scr_somatic_t | Somatic Complaints Syndrome Scale (T-score) | ASR |
| ASR_ThoughtProblems | asr_scr_thought_t | Thought Problems Syndrome Scale (T-score) | ASR |
| **ASR_AttentionProblems** | **asr_scr_attention_t** | **Attention Problems Syndrome Scale (T-score)** | **ASR** |
| **ASR_Aggression** | **asr_scr_aggressive_t** | **Aggressive Behavior Syndrome Scale (T-score)** | **ASR** |
| **ASR_RuleBreaking** | **asr_scr_rulebreak_t** | **Rule-Breaking Behavior Syndrome Scale (T-score)** | **ASR** |
| **ASR_Intrusiveness** | **asr_scr_intrusive_t** | **Intrusive Syndrome Scale (T-score)** | **ASR** |
| **ASR_Internalizing** | **asr_scr_internal_t** | **Internalizing Problems Syndrome Scale (T-score)** | **ASR** |
| **ASR_Externalizing** | **asr_scr_external_t** | **Externalizing Problems Syndrome Scale (T-score)** | **ASR** |
| **ASR_TotalProblems** | **asr_scr_totprob_t** | **Total Problems Syndrome Scale (T-score)** | **ASR** |
| **ASR_Depressive** | **asr_scr_depress_t** | **Depressive Problems DSM-5 Oriented Scale (T-score)** | **ASR** |
| ASR_AnxietyDisorder | asr_scr_anxdisord_t | Anxiety Problems DSM-5 Oriented Scale (T-score) | ASR |
| **ASR_SomaticProblems** | **asr_scr_somaticpr_t** | **Somatic Problems DSM-5 Oriented Scale (T-score)** | **ASR** |
| **ASR_Avoidant** | **asr_scr_avoidant_t** | **Avoidant Personality DSM-5 Oriented Scale (T-score)** | **ASR** |
| **ASR_ADHD** | **asr_scr_adhd_t** | **ADHD Problems DSM-5 Oriented Scale (T-score)** | **ASR** |
| **ASR_Antisocial** | **asr_scr_antisocial_t** | **Antisocial Personality DSM-5 Oriented Scale (T-score)** | **ASR** |
| ASR_Inattention | asr_scr_inattention_t | Inattention DSM-5 Oriented Scale (T-score) | ASR |
| **ASR_Hyperactive** | **asr_scr_hyperactive_t** | **Hyperactivity DSM-5 Oriented Scale (T-score)** | **ASR** |
| **ABCL_TobaccoUse** | **abcl_scr_sub_use_tobacco_t** | **Tobacco Use (T-score)** | **ABCL** |
| **ABCL_AlcoholUse** | **abcl_scr_sub_use_alcohol_t** | **Alcohol Use (T-score)** | **ABCL** |
| **ABCL_DrugUs** | **abcl_scr_sub_use_drugs_t** | **Drug Use (T-score)** | **ABCL** |
| **ABCL_SubstanceUseAvg** | **abcl_scr_sub_use_t_mean** | **Mean Substance Use Score (T-score)** | **ABCL** |
| ABCL_AdaptationFriends | abcl_scr_adapt_friends_t | Adaptive Functioning—Friends Scale (T-score) | ABCL |
| ABCL_Anxious | abcl_scr_prob_anxious_t | Anxious/Depressed Syndrome Scale (T-score) | ABCL |
| ABCL_Withdrawn | abcl_scr_prob_withdrawn_t | Withdrawn/Depressed Syndrome Scale (T-score) | ABCL |
| ABCL_SomaticComplaints | abcl_scr_prob_somatic_t | Somatic Complaints Syndrome Scale (T-score) | ABCL |
| ABCL_ThoughtProblems | abcl_scr_prob_thought_t | Thought Problems Syndrome Scale (T-score) | ABCL |
| ABCL_AttentionProblems | abcl_scr_prob_attention_t | Attention Problems Syndrome Scale (T-score) | ABCL |
| **ABCL_Aggression** | **abcl_scr_prob_aggressive_t** | **Aggressive Behavior Syndrome Scale (T-score)** | **ABCL** |
| **ABCL_RuleBreaking** | **abcl_scr_prob_rulebreak_t** | **Rule-Breaking Behavior Syndrome Scale (T-score)** | **ABCL** |
| ABCL_Intrusiveness | abcl_scr_prob_intrusive_t | Intrusive Syndrome Scale (T-score) | ABCL |
| **ABCL_Internalizing** | **abcl_scr_prob_internal_t** | **Internalizing Problems Syndrome Scale (T-score)** | **ABCL** |
| **ABCL_Externalizing** | **abcl_scr_prob_external_t** | **Externalizing Problems Syndrome Scale (T-score)** | **ABCL** |
| **ABCL_TotalProblems** | **abcl_scr_prob_total_t** | **Total Problems Syndrome Scale (T-score)** | **ABCL** |
| ABCL_CriticalProblems | abcl_scr_prob_critical_t | Critical Items Syndrome Scale (T-score) | ABCL |

Source: ABCD dataset ASR and ABCL lists.

behaviors (node `External`), followed by overall externalizing/internalizing issues reported by spouses (node `EmoIssues`). Interestingly, self-reported parental behavioral problems (node `Behavioral`) showed a weaker impact, potentially reflecting biases in self-assessments compared to external observations. Third, children's PRS also contribute to externalizing behaviors, highlighting a genetic component. However, this direct genetic effect (approximately 0.07) was notably smaller than the effects of parental alcohol misuse (0.33) and behavioral issues (0.20).

To simplify the final model, we excluded the "parental intrusive behavior" node to reduce noise. This variable functioned as a sink in our DAG—receiving input but exerting no downstream influence—making it less informative for understanding causal pathways.

When both direct and indirect paths are considered (ie, total effects in the SEM model), parental substance misuse collectively has the strongest overall impact on children's externalizing scores. Specifically, the total effect of parental alcohol/general substance use on `External` was highest (0.7674), with additional contributions from parental drug use (0.1873) and tobacco use (0.1161), resulting in a combined substance misuse effect exceeding 1.1. The next-largest total effect was observed for parental overall externalizing/internalizing issues (spouse report) at 0.4028, while the total effect for PRS was 0.1428. These total effect values represent the expected change in the child's externalizing score (in standard units) for every 1-unit increase in each predictor, capturing both direct and indirect effects. Note that these values do not reflect the total variance explained ($R^2$) by the model.

All path coefficients—both direct and indirect—were estimated using SEM derived from our causal inference model, and nearly all were highly significant ($P < 0.0005$). Full results are available in the Supplementary Materials. Overall, these findings underscore the critical role of parental alcohol and substance use, as well as externalizing/internalizing behaviors, in shaping children's externalizing problems—while also highlighting a smaller, yet meaningful, genetic influence as captured by PRS.

To further corroborate the SEM findings, we conducted two complementary regression-based analyses: (1) single-predictor ordinary least squares (OLS) regressions for each parental factor and PRS, and (2) a LASSO regression incorporating all predictors simultaneously.

## Univariate (Single-Predictor) regressions

We regressed children's externalizing scores on each parental factor and the PRS individually. Notably, *Parents Broad Behavioral & Emotional Dysregulation (Self Report)* yielded the largest effect size ($\beta = 0.456$, $R^2 = 0.208$) among the parental-behavior measures (Table 2). Other parental factors, such as *Parents Overall Externalizing/Internalizing Issues (Spouse Report)* ($\beta = 0.366$, $R^2 = 0.134$), also demonstrated significant positive associations with externalizing. Measures of parental substance use (eg, tobacco, drug, and alcohol components) were likewise significant but explained smaller portions of variance ($R^2$ typically between 0.002 and 0.011). The PRS slope ($\beta = 0.074$) was also significant, albeit with a modest $R^2$ (0.005).

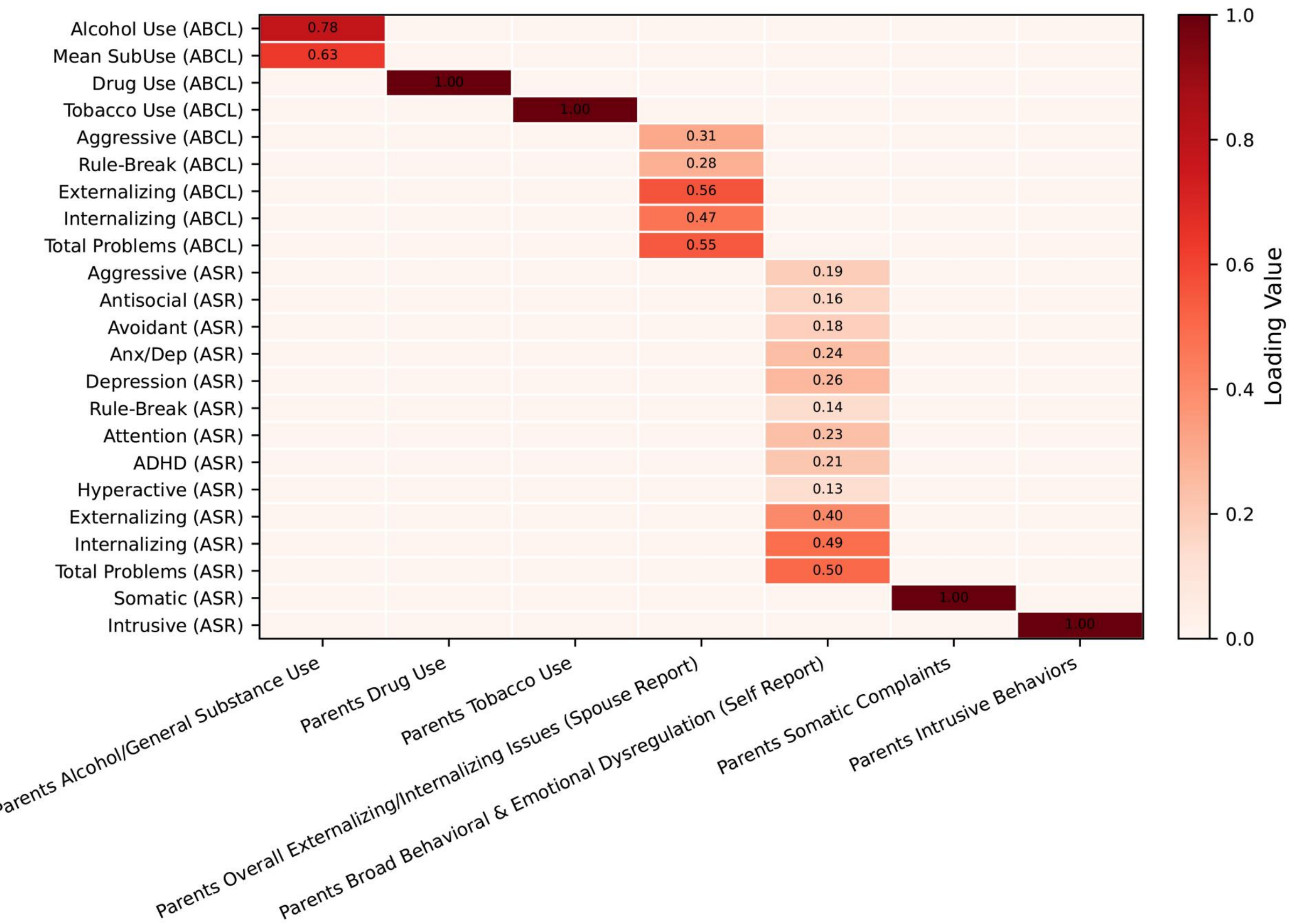

**Figure 4.** Aggregated parental factors and the loadings for all selected parental PCs.

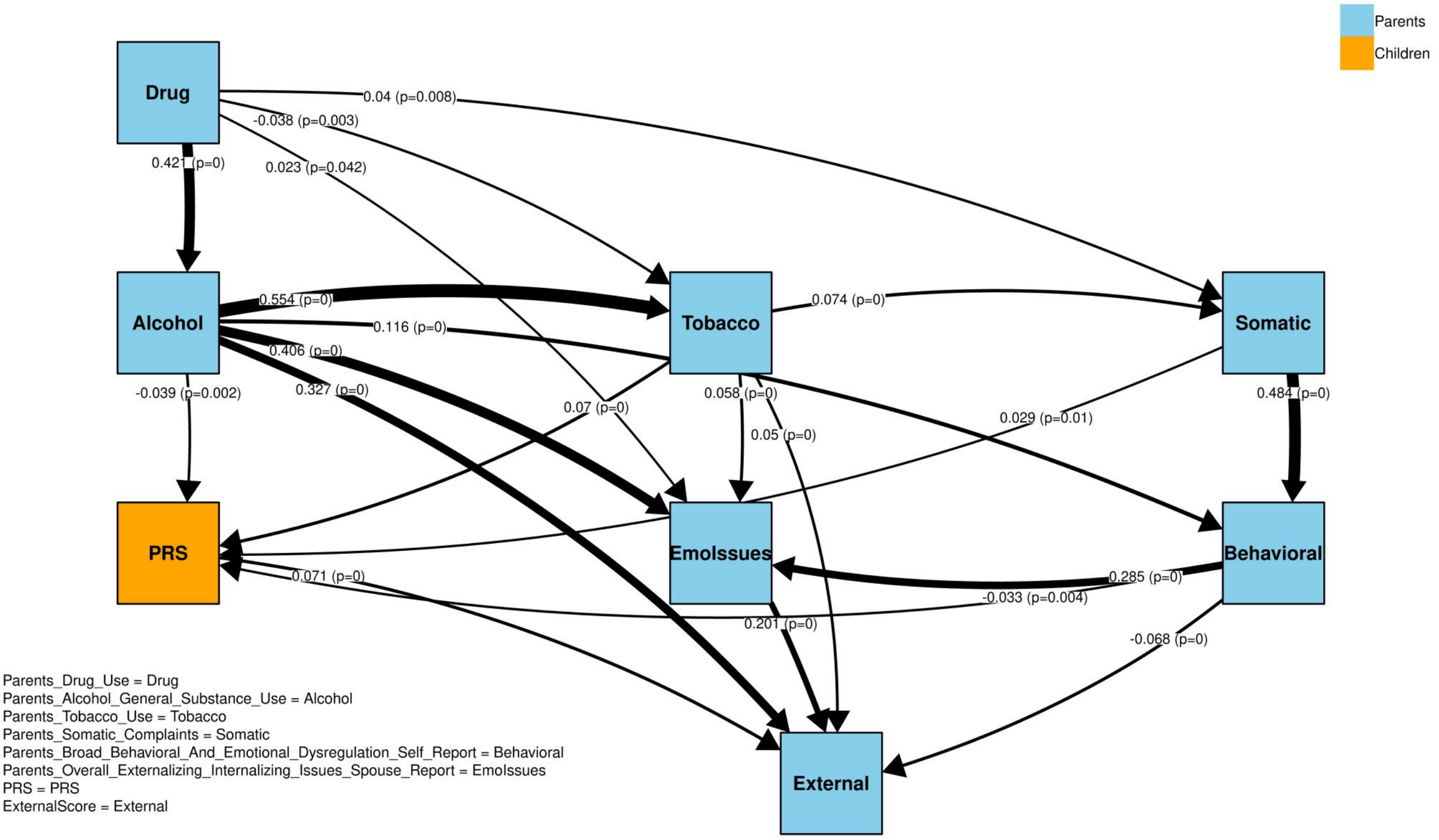

**Figure 5.** The output direct paths using SEM calculations from the causal inference model. The labels on each edge indicate the effect size, which is also represented by the edge thickness, with p-values shown in parentheses.

**Table 2.** Results from single-predictor OLS regressions.

| Predictor | Slope | *P*-value | $R^2$ |
|---|---|---|---|
| Parents broad behavioral & emotional dysregulation (self) | 0.456 | ∼0.0 | 0.208 |
| Parents overall externalizing/ internalizing issues (spouse) | 0.366 | ∼0.0 | 0.134 |
| Parents somatic complaints | 0.285 | ∼0.0 | 0.081 |
| Parents intrusive behaviors | 0.192 | ∼0.0 | 0.037 |
| Parents alcohol/general substance usex | 0.062 | 5.01e-09 | 0.004 |
| Parents tobacco use | 0.104 | 7.15e-23 | 0.011 |
| Parents drug use | 0.043 | 4.92e-05 | 0.002 |
| PRS | 0.074 | 2.40e-12 | 0.005 |

## LASSO regression

We then fitted a LASSO model (Table 3) to assess the relative importance of these predictors when considered jointly and to mitigate potential collinearity. The LASSO final solution (chosen regularization parameter $\alpha = 0.00197$) retained *Parents Broad Behavioral & Emotional Dysregulation (Self Report)* with the highest coefficient ($\beta = 0.351$), followed by *Parents Overall Externalizing/Internalizing Issues (Spouse Report)* ($\beta = 0.191$) and the PRS ($\beta = 0.075$). Smaller yet nonzero coefficients were detected for *Parents Somatic Complaints* ($\beta = 0.017$) and *Parents Tobacco Use* ($\beta = 0.027$). Interestingly, the coefficient for *Parents Intrusive Behaviors* was shrunk to zero, suggesting it did not substantially contribute once other parental variables were accounted for, which is consistent with the causal model we identified. Overall, these predictors collectively explained 23.97% of the variance in externalizing scores ($R^2 = 0.2397$).

Taken together, the single-predictor regressions highlight that each parental measure—including alcohol, tobacco, and drug use—has a statistically significant relationship with children's externalizing behaviors. However, in the multi-predictor LASSO model, parental behavioral dysregulation and parental externalizing/internalizing issues remain the dominant factors, alongside a modest contribution from the PRS. These results align closely with the SEM pathway analysis, reinforcing that (1) multiple parental traits, especially alcohol misuse and overall externalizing/internalizing issues, play central roles in shaping children's externalizing behaviors, and (2) the genetic effect indexed by PRS, while significant, is smaller in magnitude compared to these environmental influences.

## Model verification

### Bootstrap analysis

We compared our primary causal model—SEM paths (Figure 5)—with the network structure derived from the bootstrap resampling procedure (see Figure S1 and Table S3). Most SEM-derived weights fall within their corresponding bootstrap-derived confidence intervals, and the edges with the highest frequencies in the bootstrap analysis align with the strongest direct paths in our SEM. This close agreement indicates that our core causal structure is robust, while any edges with lower frequency or whose SEM weights fall outside their bootstrap intervals flag areas where further scrutiny or larger sample sizes may be needed. For instance, the direct paths from Parents Alcohol/General Substance Use, from PRS, and from Parents' Overall Externalizing/Internalizing Issues to External all remained robust across resamples, suggesting stable relationships. The highest-frequency edges in the bootstrap network correspond closely to the strongest SEM paths (eg, from Parents Substance Misuse to External or from Parents Behavioral Issues to External), reinforcing that these are central, repeatable effects in our model.

By combining SEM with a bootstrap-based DAG stability check, we highlight pathways most consistently supported by the data. Detailed edge-by-edge statistics from the bootstrap analysis are provided in Supplementary Table S9.

**Table 3.** LASSO regression coefficients.

| Predictor | Coefficient |
|---|---|
| Parents broad behavioral & emotional dysregulation (self) | 0.3510 |
| Parents overall externalizing/internalizing issues (spouse) | 0.1906 |
| PRS | 0.0750 |
| Parents somatic complaints | 0.0173 |
| Parents tobacco use | 0.0269 |
| Parents alcohol/general substance usex | −0.0582 |
| Parents drug use | −0.0272 |
| Parents intrusive behaviors | 0.0 |

### Sensitivity analysis

Our sensitivity analysis identified 15 overlapping edges (out of 28 unique edges) shared between the European-descendants-only model and the full-sample model, corresponding to a Jaccard overlap of approximately 53.6%. Specifically, approximately 71.4% of the edges observed in the European-descendants-only model were also present in the full sample, and 68.2% of the edges in the full sample were observed in the European-descendants subset. The edges common to both groups, as well as those unique to each, are visualized in Figure 6.

These findings mirror the principal patterns observed in the main analysis: parental substance misuse (particularly alcohol use) continues to show a strong influence on children's externalizing behaviors, and both parental behavioral factors and child genetic risk (PRS) contribute meaningfully to these outcomes. Similar conclusions also emerged from our structural equation modeling (SEM) and bootstrap analyses, reinforcing the robustness of our initial findings. Additional results and detailed subgroup analyze specific to the European-descendants cohort are provided in the Supplementary Information.

## Discussions and conclusions

Our findings underscore the pivotal role of parental behaviors—especially substance use and related emotional/behavioral problems—in shaping children's externalizing outcomes. By applying a data-driven causal inference framework, we uncovered both direct pathways (eg, parental alcohol and tobacco use → child externalizing score) and indirect routes mediated by broader parental dysfunction (eg, substance misuse → emotional dysregulation → child behavior). Although modest, children's PRS also contributed to externalizing behaviors, confirming a genetic predisposition component. Importantly, parental behaviors themselves reflect a blend of genetic and environmental influences. The very low correlation between PRS and parental measures ($r \approx 0.05$) indicates that each captures unique variance, supporting the value of integrating genetic scores with behavioral data.[25]

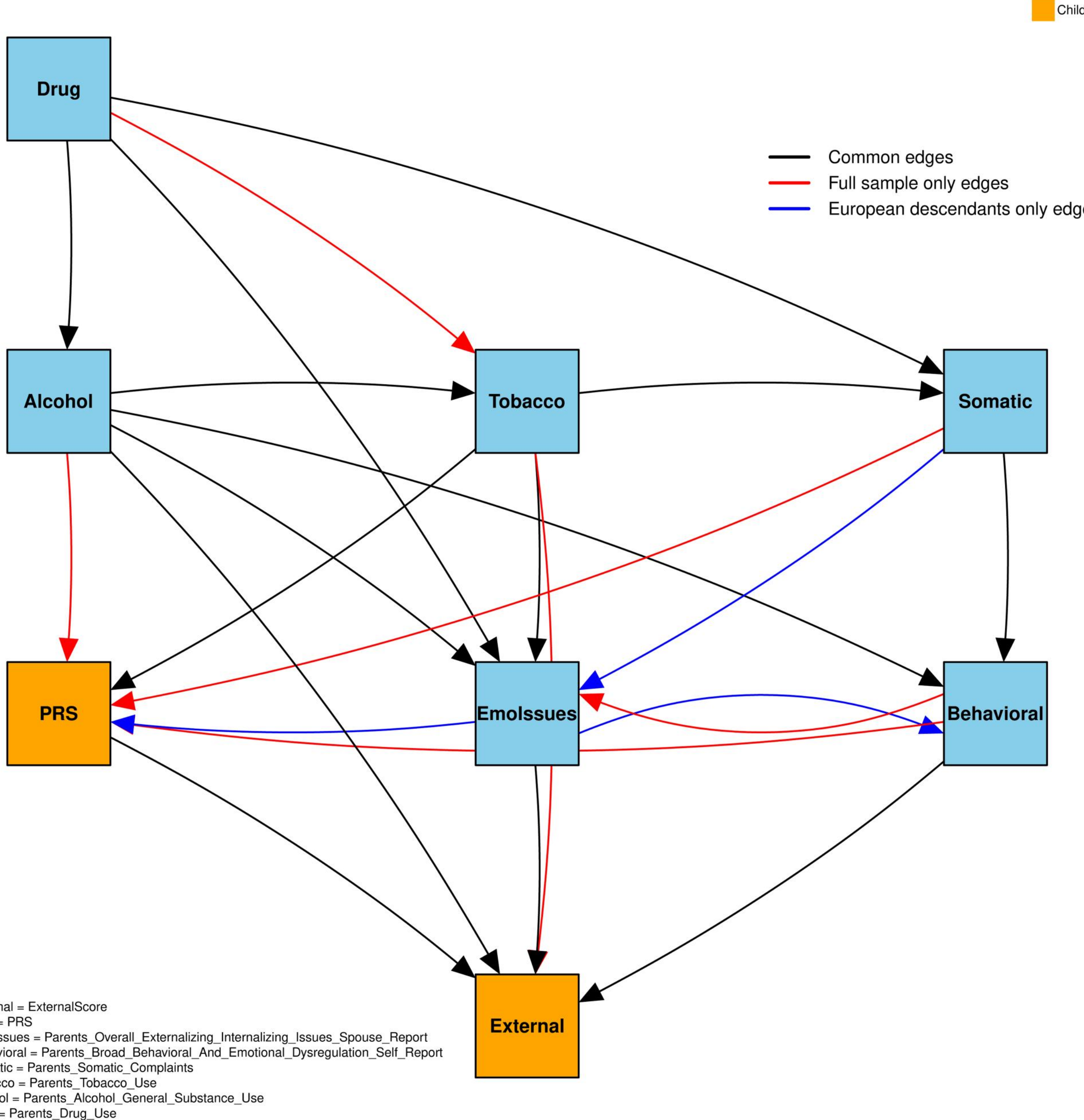

**Figure 6.** The edges that are common and unique to all samples and European descendants.

As a proof of concept, our causal model demonstrates sensitivity to modeling specifications. DAG-based causal discovery is particularly affected by correlations among variables. In our primary analysis, the inferred causal pathways generally progress from parental substance use to parental emotional dysregulation, and subsequently to children's externalizing behaviors. However, increasing the similarity threshold during clustering of parental factors—resulting in more, and more highly correlated, nodes—alters the DAG structure. Specifically, some edges involving parental substance uses and behaviors reverse direction (see Figure S2 and Table S11). The bootstrap analysis, however, reveals many of these edges to be unstable (Figure S3, Table S12). Although some of the reversed directions may be theoretically plausible, the observed ambiguity and statistical instability are more likely attributable to increased multicollinearity among nodes. These findings highlight how parameter settings and modeling constraints can influence the resulting network structure, as well as the direction and interpretation of inferred causal relationships. We also remark that while some inferred edges appear to flow from parental behaviors to children's PRS, they are not strictly causal. Rather, they reflect the fact that parental genetics influence both their own behaviors and their children's inherited genetic risk for externalizing behaviors.

From a public health and intervention standpoint, these findings offer several actionable insights. First, the identification of parental substance use and emotional dysregulation as upstream causal factors highlights the value of family-centered interventions. Clinicians and social workers may consider prioritizing support programs targeting parental mental health and substance misuse as part of preventative strategies for youth at risk of externalizing behavior. Second, the ability to identify children at genetic risk—even with modest PRS accuracy—opens avenues for early screening and stratified support, especially when combined with family-based risk profiles. Educators and school counselors may benefit from greater awareness of these multilayered risks to guide behavioral interventions. Finally, policymakers aiming to reduce adolescent behavioral health problems may find support for integrated prevention models that address both child vulnerability and parental wellness.

Our findings should be interpreted in light of the following limitations. First, our use of a DAG-based HillClimb causal model assumes acyclic relationships and therefore cannot capture feedback loops or bidirectional effects. For example, although prior research has linked parental intrusiveness to children's externalizing behavior, including possible bidirectional relationships,[26,27] our final model did not identify such a pathway—possibly due to confounding in other studies or constraints inherent to our modeling approach. Acyclic modeling also precludes us from simultaneously capturing parent-to-child and potential child-to-parent effects, such as how children's genetic predispositions or behaviors might influence parental responses. Second, the PRS reflects only a small portion of heritable risk, as it is derived from GWAS summary statistics and does not capture the full genetic architecture of externalizing behaviors in the ABCD cohort. The necessary coordinate liftover step between genome builds (GRCh37 to GRCh38) also resulted in SNP loss, potentially limiting PRS accuracy. Third, although the ABCD dataset is large and diverse, the findings may not generalize to non-U.S. populations or culturally distinct family environments.

In conclusion, this study demonstrates the value of combining causal inference techniques with behavioral and genetic data to uncover potential mechanisms driving externalizing behaviors in youth. The substantial impact of parental behavioral factors suggests that modifying these risks could have a meaningful effect on children's outcomes. Future work should build on this foundation by integrating more comprehensive environmental metrics (eg, neighborhood adversity, peer influences)[28] and improving genetic predictors to develop more holistic intervention strategies.

## Author contributions

Mengman Wei: Conceptualization; Methodology; Software; Formal analysis; Data curation; Visualization; Writing – original draft.

Qian Peng: Conceptualization; Validation; Supervision; Project administration; Funding acquisition; Writing – review & editing.

All authors read and approved the final manuscript.

## Supplementary material

Supplementary material is available at *Journal of the American Medical Informatics Association* online.

## Funding

This work was supported by the National Institutes of Health (NIH), National Institute on Drug Abuse (NIDA) under award DP1DA054373 (to Qian Peng). The funder had no role in the study design; data collection, analysis, or interpretation; manuscript writing; or the decision to submit for publication. The content is solely the responsibility of the authors and does not necessarily represent the official views of the NIH.

## Conflicts of Interest

The authors declare no competing interests.

## Data availability

**Code.** The analysis code and scripts used in this study are freely available at the following GitHub repository: https://github.com/mw742/Hillclimb-Causal-Inference.git

**Data.** This study uses data from the Adolescent Brain Cognitive Development (ABCD) Study (abcdstudy.org), held in the NIMH Data Archive (NDA). The ABCD data version used was 5.1. The study is supported by the National Institutes of Health (NIH) and additional federal partners under multiple award numbers, including `U01DA041048` and `U01DA050987`. The full list of funders is available at abcdstudy.org/federal-partners.html.